\documentclass[12pt]{article}
\usepackage{graphicx}
\usepackage[utf8]{inputenc}
\usepackage{amsmath}
\usepackage{lineno}

\newcommand\pubnumber{NuPhys2016-Migenda}
\newcommand\pubdate{\today}

\def\Title#1{\begin{center} {\Large #1 } \end{center}}
\def\Author#1{\begin{center}{ \sc #1} \end{center}}
\def\Address#1{\begin{center}{ \it #1} \end{center}}

\newcommand\pubblock{\rightline{\begin{tabular}{l} \pubnumber\\
         \pubdate  \end{tabular}}}
\newenvironment{Abstract}{\begin{quotation}  }{\end{quotation}}
\newenvironment{Presented}{\begin{quotation} \begin{center} 
             PRESENTED AT\end{center}\bigskip 
      \begin{center}\begin{large}}{\end{large}\end{center} \end{quotation}}

\def\beq{\begin{equation}}
\def\eeq#1{\label{#1}\end{equation}}
\def\eeqn{\end{equation}}

\def\beqa{\begin{eqnarray}}
\def\eeqa#1{\label{#1}\end{eqnarray}}
\def\eeqan{\end{eqnarray}}

\let\bar=\overbar

\def\Dslash{\not{\hbox{\kern-4pt $D$}}}
\def\dslash{\not{\hbox{\kern-2pt $\del$}}}

\def\msb{{\bar{\ssstyle M \kern -1pt S}}}

\begin{document}
\begin{titlepage}
\pubblock

\vfill
\Title{The Hyper-Kamiokande Experiment: Overview \& Status}
\vfill
\Author{Jost Migenda for the Hyper-Kamiokande proto-collaboration}
\Address{University of Sheffield, Department of Physics \& Astronomy, Sheffield S3~7RH, UK}
\vfill
\begin{Abstract}
The Hyper-Kamiokande (HK) experiment centres around a proposed next-generation underground water Cherenkov detector that will be nearly 20 times larger than the highly successful Super-Kamiokande experiment and use significantly improved photodetectors with the same 40\,\% photocoverage.

HK will increase existing sensitivity to proton decay by an order of magnitude, and it will study neutrinos from various sources, including atmospheric neutrinos, solar neutrinos, and supernova neutrinos. In addition to operating as a standalone experiment, HK will serve as the far detector of a long-baseline neutrino experiment using the upgraded J-PARC neutrino beam, enhancing searches for lepton-sector CP violation.

This poster presents recent developments and the current status of the experiment. It provides an overview of the project, including ongoing R\&D efforts and upgrades to both the beam and the near detector suite. The expected physics reach, showcased in the recently published design report, will also be featured.

[Nb: This contribution to the NuPhys2016 proceedings focuses on photo\-sensor development and supernova neutrinos. Other physics to\-pics, including neutrino oscillations and nucleon decay, are discussed in a se\-pa\-rate contribution to these proceedings.]
\end{Abstract}
\vfill
\begin{Presented}
NuPhys2016, Prospects in Neutrino Physics\\
Barbican Centre, London, UK, December 12--14, 2015
\end{Presented}
\vfill
\end{titlepage}
\def\thefootnote{\fnsymbol{footnote}}
\setcounter{footnote}{0}

\section{Introduction}
Hyper-Kamiokande (HK,~\cite{PublicDR}) is a proposed next-generation water Cherenkov detector, whose broad physics programme covers many areas of particle and astroparticle physics.
It will increase existing sensitivity to nucleon decay by an order of magnitude, and it will study neutrinos from various sources, including atmospheric neutrinos, solar neutrinos, supernova neutrinos and annihilating dark matter. In addition to operating as a standalone experiment, HK will serve as the far detector of a long-baseline neutrino experiment (T2HK) using the upgraded J-PARC neutrino beam, enhancing searches for lepton-sector CP violation and enabling precision measurements of several other neutrino oscillation parameters.

In section~\ref{sec:hk}, we give an overview over the experiment.
We discuss photosensor development in section~\ref{sec:pmt} and the expected physics reach in the area of supernova neutrinos in section~\ref{sec:sn}.
Other physics topics, in particular neutrino oscillations and nucleon decay, are discussed in a separate contribution to these proceedings~\cite{NuPhys2016}.

\section{The Hyper-Kamiokande Experiment}\label{sec:hk}

\subsection{Overview}

\begin{figure}[htb]
\centering
\includegraphics[width=28pc]{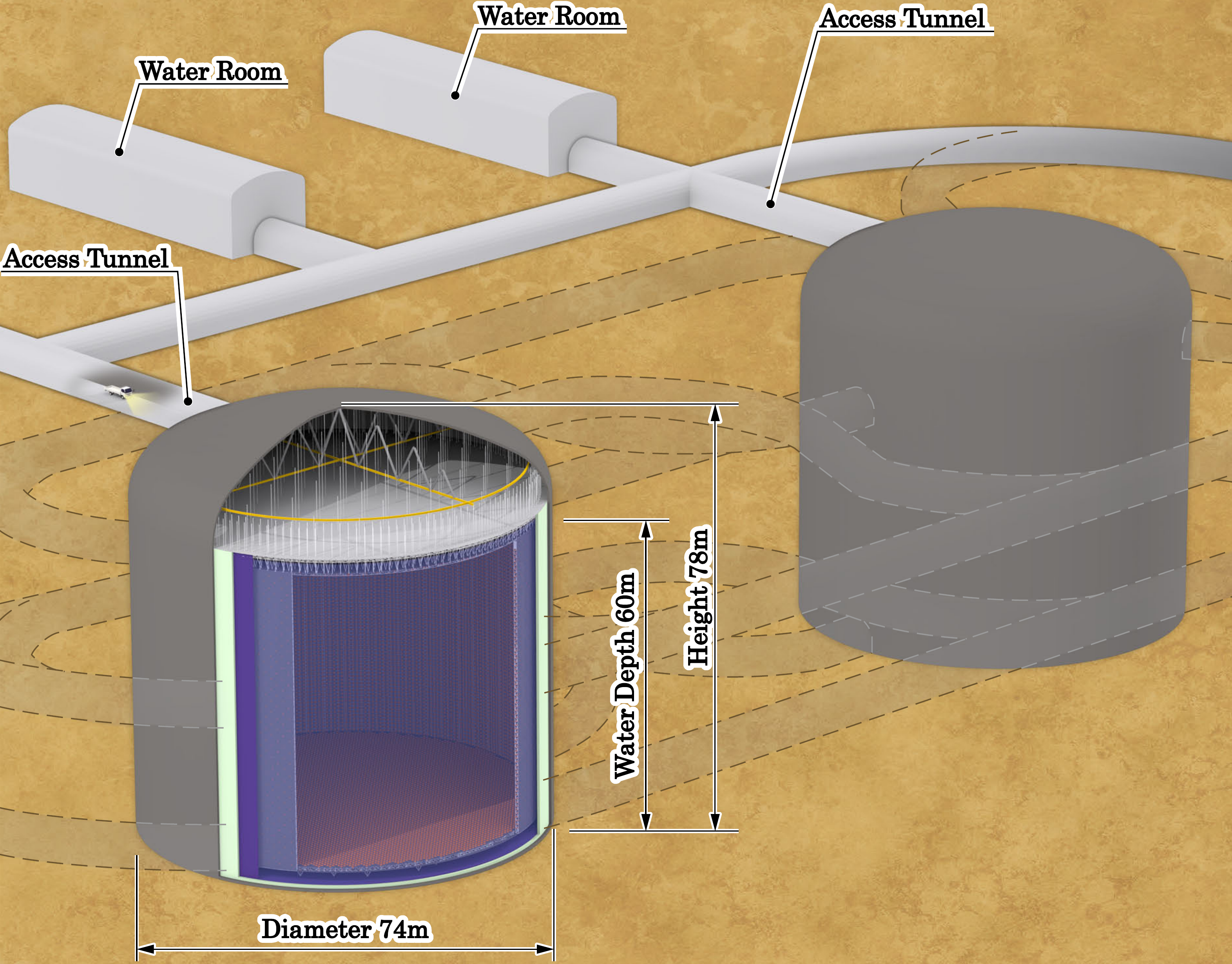}
\caption{Drawing of the Hyper-Kamiokande detector and infrastructure.}
\label{fig:hk}
\end{figure}

HK is based on the proven technology of (Super-)Kamiokande. Its much higher detector volume and additional improvements in key areas like photosensors and near/intermediate detectors make it a straightforward yet powerful extension of the very successful (Nobel Prizes~2002~and~2015) Japan-based neutrino programme.

The detector will be located about 8\,km south of Super-Kamiokande (SK) in the Tochibora mine with an overburden of 1750\,m.w.e. As shown in fig.~\ref{fig:hk}, it will consist of two cylindrical tanks (each 60\,m high and 74\,m in diameter) and have a total (fiducial) mass of 0.52 (0.37) Mton, making it 10 (17) times larger than its predecessor.
It will use 40,000 photomultiplier tubes (PMTs) per tank to reach the same 40\,\% photocoverage, and benefit from newly designed high-efficiency PMTs.

\subsection{Recent Progress}
The HK proto-collaboration was formed in January 2015 and has since grown to about 300 people from 73 institutions and 15 countries.
In 2016, a design report presenting an optimized detector design was published~\cite{PublicDR}.

Construction for the first tank is expected to start in 2018, with data-taking starting in 2026.
In September 2016, the T2K collaboration published a proposal for an extended run of their experiment~\cite{T2K2}, which would enable a seamless transition to T2HK. An upgrade of the J-PARC beamline to 750\,kW power is already ongoing, while additional upgrades to reach 1.3\,MW are planned before the start of HK. 

In the baseline design described in the design report, the second tank would get constructed next to the first one and start data-taking in 2032. As an alternative, the possibility of building the second tank in Korea was explored in a white paper published in November 2016~\cite{T2HKK}. At a longer baseline of 1000--1300\,km, that tank would be able to observe the second oscillation maximum, where the effect of a non-zero $\delta_\text{CP}$ would be increased.
The proposed detector locations offer a higher overburden (and thus lower spallation backgrounds) than the Japanese HK site, which would increase sensitivity to low-energy physics like solar or supernova relic neutrinos.

\section{Photosensor Development}\label{sec:pmt}

\begin{figure}[htb]
\hspace{-0.5pc}\begin{minipage}{15pc}
\includegraphics[width=17pc]{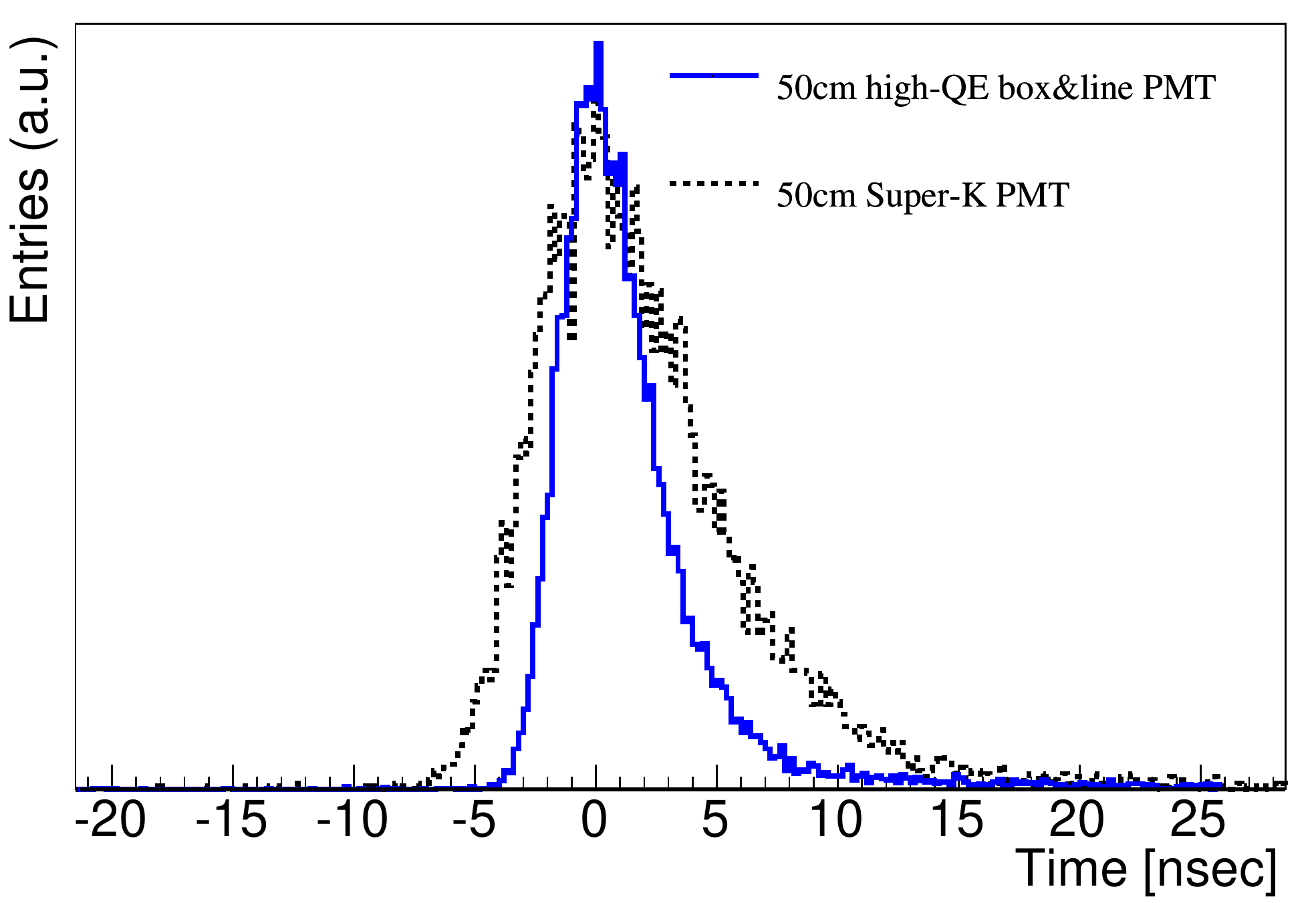}
\end{minipage}\hspace{2pc}
\begin{minipage}{18pc}
\includegraphics[width=20pc]{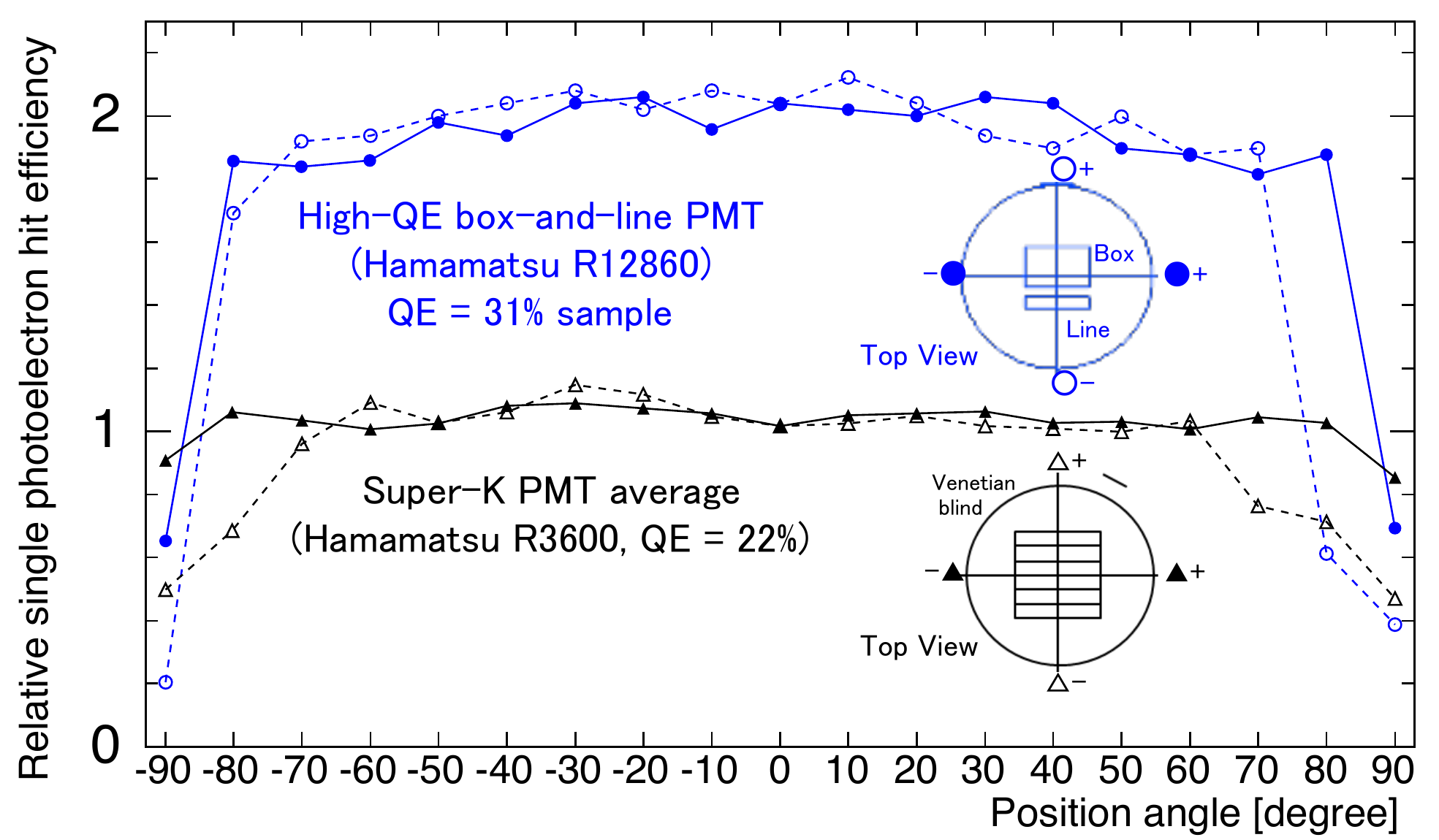}
\end{minipage} 
\caption{Single p.e. transit time (left) and relative detection efficiency (right) in the HQE B\&L PMT developed for HK (blue) and the PMT used in SK (black).}
\label{fig:pmt}
\end{figure}

A new 50 cm PMT model, the Hamamatsu R12860-HQE, was developed for HK. It is based on Hamamatsu’s R3600 PMT used in SK, but includes a box-and-line dynode and several other improvements. As a result, this new model offers better timing resolution and a detection efficiency that is two times as high due to improvements to both quantum efficiency and collection efficiency (see fig.~\ref{fig:pmt}). Work to reduce the dark noise rate and design new PMT covers for pressure resistance is currently ongoing.

In addition to this baseline design, R\&D on alternative photosensor options like hybrid photo-detectors, LAPPDs and multi-PMT modules is ongoing.

\section{Supernova Neutrinos}\label{sec:sn}
\subsection{Galactic Supernova}

For a galactic supernova at a fiducial distance of $10\,$kpc, HK will detect $\mathcal{O}(10^5)$ neutrinos within about 10\,s.
This high event rate enables HK to resolve fast time variations of the event rate, which could give us information on properties of the progenitor (like its rotation) or on details of the supernova explosion mechanism like the roles of turbulence, convection and the standing accretion shock instability, SASI, on which there is significant disagreement between different computer simulations~\cite{Janka2016}.

\subsection{Supernovae in Neighbouring Galaxies}
Due to its large volume, HK would be sensitive to supernova bursts in nearby galaxies as well, observing $\mathcal{O}(5000)$ events from a SN1987a-like supernova in the Large Magellanic Cloud or $\mathcal{O}(20)$ events from a supernova in the Andromeda galaxy.

\begin{figure}[htb]
\hspace{-0.0pc}\begin{minipage}{16.5pc}
\includegraphics[width=16.5pc]{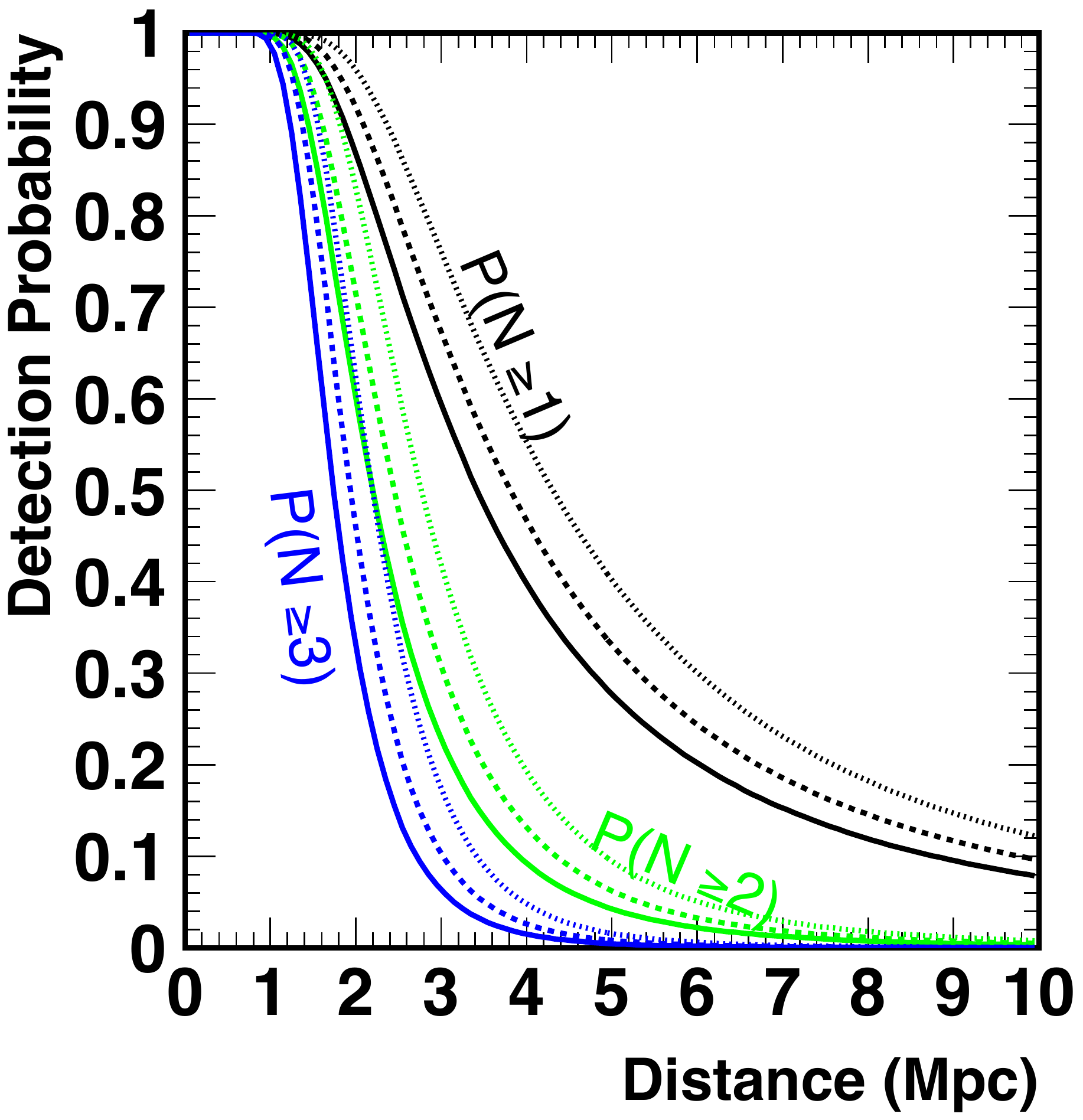}
\end{minipage}\hspace{2pc}
\begin{minipage}{16.5pc}
\includegraphics[width=16.5pc]{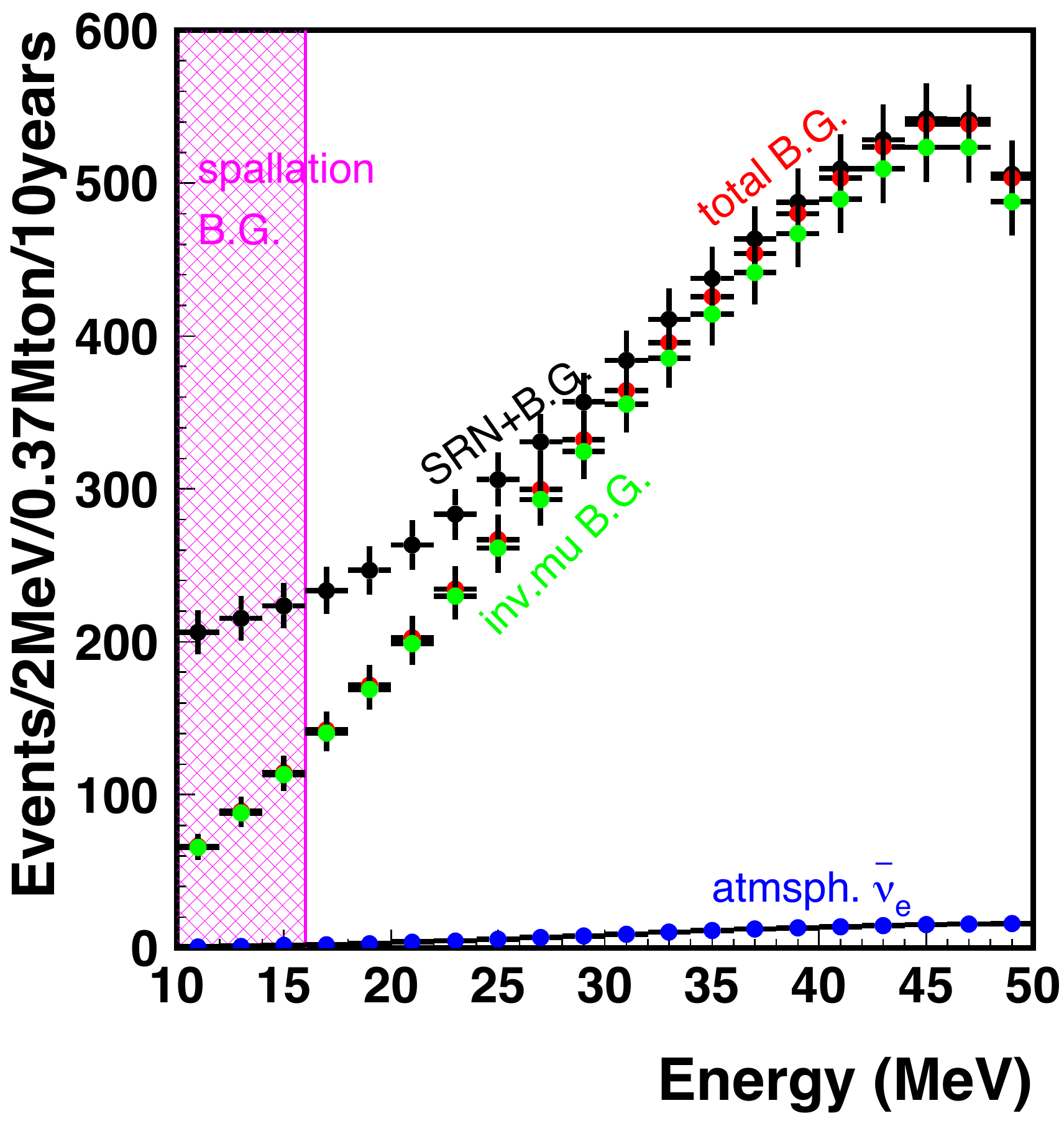}
\end{minipage} 
\caption{Left: Detection probability of supernova neutrinos versus distance at HK. Solid, dotted and dashed lines are for no oscillation, normal hierarchy and inverted hierarchy, respectively. Right: Expected spectrum of the SRN signal in HK after 10 years, including backgrounds.}
\label{fig:supernova}
\end{figure}

Using strict timing coincidence with an external trigger, like a gravitational wave signal in LIGO or the nearby KAGRA, HK could even be sensitive to single supernova neutrino events. For supernovae at up to 4\,Mpc distance, which are expected to happen every 3--4 years on average, HK has a 50\,\% or greater chance of detecting at least one event (see fig.~\ref{fig:supernova}).

\subsection{Supernova Relic Neutrinos}

The total neutrino flux from all remote supernovae in the history of the universe is known as supernova relic neutrinos (SRN). While these neutrinos cannot be traced back to a specific supernova, they deliver information on the average spectrum of supernova neutrinos and could enable a first measurement of the rate of failed (optically dark) supernovae, which are the origin of stellar mass black holes. The SRN measurement is therefore complementary to observations of nearby supernovae.

At HK, SRN are observable in an energy window of 16--30\,MeV shown in fig.~\ref{fig:supernova}, which is bounded by cosmic-ray induced spallation backgrounds at lower energies and invisible muon background from atmospheric neutrinos at higher energies. Within 10 years, about 100 SRN events are expected at HK, corresponding to an observation of SRN with 4.8\,$\sigma$ significance.


\begin{thebibliography}{5}

\bibitem{PublicDR}
  K.~Abe {\it et al.} [Hyper-Kamiokande Proto-Collaboration],
  KEK-Preprint-2016-21, ICRR-Report-701-2016-1.

\bibitem{NuPhys2016}
  See contribution by M.~Yokoyama in these proceedings.

\bibitem{T2K2}
  K.~Abe {\it et al.} [Hyper-Kamiokande Proto-Collaboration],
  arXiv:1609.04111.

\bibitem{T2HKK}
  K.~Abe {\it et al.} [Hyper-Kamiokande Proto-Collaboration],
  arXiv:1611.06118.

\bibitem{Janka2016}
  H.-Th.~Janka, T.~Melson and A.~Summa,
  Annu.~Rev.~Nucl.~Part.~Sci.~\textbf{66} (2016) p.\,341--375.
  arXiv:1602.05576.

\end{thebibliography}
\end{document}